\documentclass[a4paper,12pt]{article}
\usepackage{graphicx}
\usepackage{amsmath}
\usepackage{amssymb}
\usepackage{setspace}
\usepackage{ulem}
\usepackage{subfigure} 
\usepackage{setspace}

\begin{document}
\begin{titlepage}
\null
\begin{flushright}
March, 2017
\end{flushright}

\vskip 1.8cm
\begin{center}

  {\Large \bf Atiyah-Manton Construction of\\
\vspace{0.5cm}
Skyrmions in Eight Dimensions}

\vskip 1.8cm
\normalsize

  {\bf Atsushi Nakamula\footnote{nakamula(at)sci.kitasato-u.ac.jp},
Shin Sasaki\footnote{shin-s(at)kitasato-u.ac.jp} and Koki Takesue\footnote{ktakesue(at)sci.kitasato-u.ac.jp}}

\vskip 0.5cm

  { \it
  Department of Physics \\
  Kitasato University \\
  Sagamihara 252-0373, Japan
  }

\vskip 2cm

\begin{abstract}
We show that the eight-dimensional instanton solution, which satisfies the self-duality
 equation $F \wedge F = *_8 F \wedge F$, realizes the static Skyrmion
 configuration in eight dimensions through the Atiyah-Manton construction.
The relevant energy functional of the Skyrme field is obtained by the
 formalism developed by Sutcliffe.
By comparing the Skyrmion solution associated with the extreme of the energy, with the
Atiyah-Manton solution constructed by the instantons,
we find that they agree with high accuracy.
This is a higher-dimensional analogue of the Atiyah-Manton construction
 of Skyrmions in four dimensions.
Our result indicates that the instanton/Skyrmion correspondence seems to
 be an universal property in $4k \ (k=1, 2, \ldots)$ dimensions.

\end{abstract}
\end{center}
\end{titlepage}

\newpage
\tableofcontents
\section{Introduction}
The Skyrme model \cite{Skyrme:1962vh} is a model for pions in the
low-energy effective theory of QCD.
The model is a four-dimensional non-linear sigma model whose target space is
$S^3 \sim SU(2)$, and composed of the fourth order derivative term in addition to the canonical kinetic term.
The fourth order derivative term guarantees the stability of solitons of co-dimension three, which are called Skyrmions.
The Skyrmions are characterized by the homotopy class $\pi_3 (SU(2)) = \mathbb{Z}$
and they are regarded as Baryons.
The energy functional of the Skyrme model has the Bogomol'nyi
bound given by the topological charge associated with the homotopy.
This topological charge is identified with the Baryon number.
However, no analytic solutions that saturate the lower bound of the energy have been found so far
\footnote{
This is not the case for Skyrme models in curved spaces. For example, see \cite{Manton:1986pz,Manton:1987xt,Canfora:2014aia} and references therein.
}.
There have only been obtained the numerical solutions of Skyrmions, which indeed exceed the
energy bound.
This reflects the fact that the original four dimensional Skyrme model does not have the BPS property.

Finding proper solutions of Skyrmions is a long standing problem.
There are several directions to construct solutions.
For example, the rational map ansatz provides a good approximation to
the Skyrmion solutions \cite{Houghton:1997kg}.
This includes solutions corresponding to higher Baryon numbers.
Although they can not saturate the energy bound, the rational map
solutions have close energies to the normalized Baryon charges.
Alternatively, there is another promising approach to Skyrmions known as the Atiyah-Manton
construction \cite{Atiyah:1989dq}.
Atiyah and Manton pointed out that the holonomy of
the Yang-Mills instantons in the four-dimensional Euclid space
\footnote{
The case for the curved spaces was discussed in \cite{Manton:1990gr}.
}
 gives
a well approximated static Skyrmion solutions.
Although, the origin of this approximation is not transparent,
a physical interpretation to the Atiyah-Manton construction of Skyrmions
was discussed in \cite{Eto:2005cc, Hata:2007mb}.

Even though the Skyrmion solutions are well-approximated by instantons, they
never saturate the Bogomol'nyi bound of the energy.
In order to understand the obscure connection between the
Yang-Mills instantons and Skyrmions, we need further penetrating analysis.
In this context, in \cite{Sutcliffe:2010et}, 
inspired by a holographic QCD model \cite{Sakai:2004cn},
it is proposed a systematic derivation of the energy functional for the static Skyrme
field from the Yang-Mills action in four dimensions .
In the derivation, the introduction of the tower of mesons originated from the Kaluza-Klein-like expansion modes in higher
dimensions makes the Atiyah-Manton solution have closer energy to the
bound \cite{Sutcliffe:2015sta}.
Therefore, including the higher expansion modes in the Atiyah-Manton solution leads to
the better approximation to the Skyrmions.
Moreover, this relation is generalized to lower dimensions.
For example, an analogue of the Atiyah-Manton construction in two dimensions
is proposed \cite{Sutcliffe:1992he, Stratopoulos:1992hq} where the
sine-Gordon soliton solution in one dimensions is well-approximated by the $\mathbb{C}P^1$-lump --
the two-dimensional instantons.
These facts remarkably suggest that there is a deep correspondence between instantons 
or solitons and Skyrmion-like objects in various dimensions.

The instantons in four dimensions satisfy the self-duality
equation $F = *_4 F$. Here $F$ is the field strength 2-form of the gauge
field and $*_d$ is the Hodge dual operator in $d$ dimensions.
A natural higher-dimensional generalization of instantons is a
solution to the self-duality equations in $d = 4k$ dimensions $F^k =
*_{4k} F^k$ where $F^k$ is the $k$ wedge products of $F$.
The $k=1$ case corresponds to the ordinary instantons in four-dimensions while the
$k\ge 2$ cases are their generalization.
The first non-trivial example is the $k=2$ case, namely, the self-dual instantons in eight dimensions. 
This was studied so far from various
viewpoints \cite{Grossman:1984pi, Tchrakian:1984gq}.
On the other hand, it is possible to consider higher-dimensional
generalizations of Skyrmions \cite{Nitta:2012rq}.

In this paper we study the relation between instantons and Skyrmions in
higher dimensions. In particular, we focus on the eight-dimensional
self-dual instantons that satisfy $F \wedge F = *_8 F \wedge F$.
The self-duality relation is obtained by the Bogomol'nyi completion of
the quartic Yang-Mills action in eight dimensions.
We will derive the energy functional for the static Skyrme field from the quartic
Yang-Mills action  by the reduction procedure developed by Sutcliffe
\cite{Sutcliffe:2010et}. The Derrick's theorem indicates that the model
admits static soliton solutions which we call the eight-dimensional
Skyrmions. We will find the numerical solution of the above mentioned Skyrmion.
We will then calculate a field configuration through
the Atiyah-Manton construction applied to the eight-dimensional instanton
and find that this gives a good approximation to the numerical solution
of the Skyrmion.
Our results strongly suggest that the instanton/Skyrmion
correspondence holds even in $4k$ dimensions and this relation
is an universal property.

The organization of this paper is as follows.
In section 2, we give a brief overview of the prescription by Sutcliffe
in four dimensions.
We then derive the energy functional for the static Skyrme field in
eight dimensions from the quartic Yang-Mills action.
In section 3, we perform the numerical analysis to solve the equation of motion.
We find a spherically symmetric solution to the Skyrmions in eight dimensions.
We then calculate the holonomy associated with the self-dual
instanton in eight dimensions and construct the Atiyah-Manton solution.
We will show a good agreement between the numerical solution and the Atiyah-Manton solution.
Section 4 is an analysis of higher dimensional generalizations.
Section 5 is devoted to the conclusion and discussions.
The detail derivation of the energy functional for the static Skyrme
field with hedgehog ansatz in eight dimensions is shown in appendix.

\section{Eight-dimensional Skyrme model from quartic Yang-Mills theory}
In this section we introduce a Skyrme model in eight dimensions
following the formalism developed by Sutcliffe \cite{Sutcliffe:2010et}.
Before going to the eight-dimensional analysis, we give an overview of
the derivation for the ordinary Skyrme model in four dimensions.

\subsection{Overview of the Sutcliffe's truncation in four dimensions}
The four-dimensional energy functional for static fields\footnote{We
sometimes call this the three-dimensional action in Euclid space.} of the Skyrme model is obtained by a
reduction of the usual quadratic Yang-Mills action in the four-dimensional
Euclidean space.
The action is
\begin{align}
S = - \frac{1}{2 \kappa g^2} \int \! \text{Tr} [*_4 F \wedge F]
= - \frac{1}{4 \kappa g^2}
\int \! d^4 x \
\text{Tr} [ F_{mn} F^{mn}].
\label{eq:4dYM}
\end{align}
Here $F = \frac{1}{2!} F_{mn} dx^m \wedge dx^n, \ (m,n = 1,\ldots,4)$ is
the gauge field strength 2-form.
The component is given by
$F_{mn} = \partial_m A_n - \partial_n A_m + [A_m, A_n]$.
The gauge field $A_m$ is in the adjoint representation of a gauge group $G$
and it is expanded by the generators $T^a \ (a = 1, \ldots \text{dim}
\mathcal{G})$.
Here $\mathcal{G}$ is the Lie algebra associated with $G$ and
$\kappa$ is the normalization constant for the generators $\text{Tr}
[T^a T^b] = \kappa \delta^{ab}$.
Here $g$ is the gauge coupling constant.
Making the action \eqref{eq:4dYM} be the completely square form
results in the Bogomol'nyi-Prasad-Sommerfield (BPS) self-duality
equation $F = *_4 F$ whose solutions are called instantons.
Since the Yang-Mills action \eqref{eq:4dYM} has the scale
invariance, instanton solutions that saturate the Bogomol'nyi bound
have a size modulus.

It is proposed in \cite{Sutcliffe:2010et} that a holography-inspired
reduction of the four-dimensional Yang-Mills action \eqref{eq:4dYM}
provides the energy functional for the static Skyrme field.
Following the prescription in \cite{Sutcliffe:2010et}, we first
decompose the four-dimensional Euclidean space 
into the three-dimensional physical space and a ``fictious'' direction
: $x^m = (x^i, x^4)$ where $i = 1, \ldots, 3$.
We then expand the four-dimensional gauge field $A_m (x^i, x^4)$ in the
{\it infinite line} along the $x^4$-direction by a complete orthonormal basis
with the square integrable function.
A suitable basis with the boundary condition
$A_i (x^i, x^4) \to 0$ as $x^4 \to \infty$ is the Hermite function
$\psi_n (z) = \frac{(-1)^n}{\sqrt{n! 2^n \sqrt{\pi}}} e^{\frac{1}{2}
z^2} \frac{d^n}{dz^n} e^{-z^2}$.
Then we have an expansion,
\begin{align}
A_m (x^i,x^4) = \sum_{n=0}^{\infty} \mathcal{A}^{(n)}_m (x^i) \psi_n (x^4),
\end{align}
where $\mathcal{A}_m^{(n)} (x^i)$ are expansion coefficients, which will
be determined later.
Next, we perform the gauge transformation by which the component $A_4$ is set to be
zero. By this gauge transformation, the components of the gauge field
$A_i$ is transformed as
\begin{align}
A_i \ \longrightarrow \ \hat{g} A_i \hat{g}^{-1} + \hat{g} \partial_i \hat{g}^{-1},
\end{align}
where the gauge parameter $\hat{g}$ is given by
\begin{align}
\hat{g} (x^i, x^4) = -\text{P} \exp
\left[
\int^{x^4}_{-\infty} \! d \xi \ A_4 (x^i, \xi)
\right].
\end{align}
Here the symbol $\text{P}$ stands for the path-ordering.
The asymptotic behavior of the Hermite function $\psi_n (\infty) = 0$
and the boundary condition $A_i (x^i, \infty) = 0$ determines the
gauge field $A_i (x^i,x^4)$ in the gauge $A_4 = 0$. This is given by
\cite{Sutcliffe:2010et},
\begin{align}
A_i (x^i, x^4) = u_i (x^i) \psi_{+} (x^4) + \sum_{n=0}^{\infty} W_i^n
 (x^i) \psi_n (x^4),
\end{align}
where $\psi_+ (z) = \frac{1}{2} + \frac{1}{2} \text{erf}
(z/\sqrt{2})$ and the error function is defined by $\text{erf} (z) = \frac{2}{\sqrt{\pi}}
\int^z_0 \! d \xi \ e^{-\xi^2}$.
The gauge field is decomposed into the ``zero-mode'' $u (x^i)$:
\begin{align}
u_i (x^i) = U\partial_i U^{-1}, \qquad U(x^i) = 
\hat{g} (x^i, x^4 = \infty),
\end{align}
and the infinite tower of the vector fields $W_i^n (x^i)$.
The zero-mode $u(x^i)$ is identified with the Skyrme field while the
higher modes $W_i^n (x^i)$ can be interpreted as ``vector mesons''.
This analysis is completely parallel to the Kaluza-Klein reduction in
which a field is expanded by the Fourier modes $e^{in x^4/2\pi R}$ along the
compact circle $x^4 \sim x^4 + 2 \pi R$.
Note that the expansion along an infinite line enable us to realize
the Skyrme field $U$ by the holonomy of the gauge field:
\begin{align}
U(x^i) = -\text{P} \exp
\left[
\int^{\infty}_{- \infty} \! dx^4 \ A_4 (x^i,x^4)
\right].
\end{align}
Although it is possible to compute $W_i^n$,
let us focus on the leading approximation, {\it i.e.} we neglect all the vector meson modes
and focus only on the Skyrme field $U(x^i)$.
We call this the Sutcliffe's truncation.
Then, in the gauge $A_4 = 0$, we have the following decomposition of the
gauge field strength:
\begin{align}
F_{i4} =& \ U\partial_i U^{-1} \partial_4\psi_+ (x^4) =  R_i
 \frac{\psi_0 (x^4)}{\sqrt{2} \pi^{\frac{1}{4}}}, \notag \\
F_{ij} =& \ [R_i, R_j] \psi_+ (x^4) (\psi_{+} (x^4) - 1),
\label{eq:gauge_Skyrme}
\end{align}
where $R_i = U\partial_i U^{-1}$ is interpreted as the right current.

\begin{figure}[tb]
\begin{center}
\subfigure[The profile for 4d Skyrmion.]
{
\includegraphics[scale=.6]{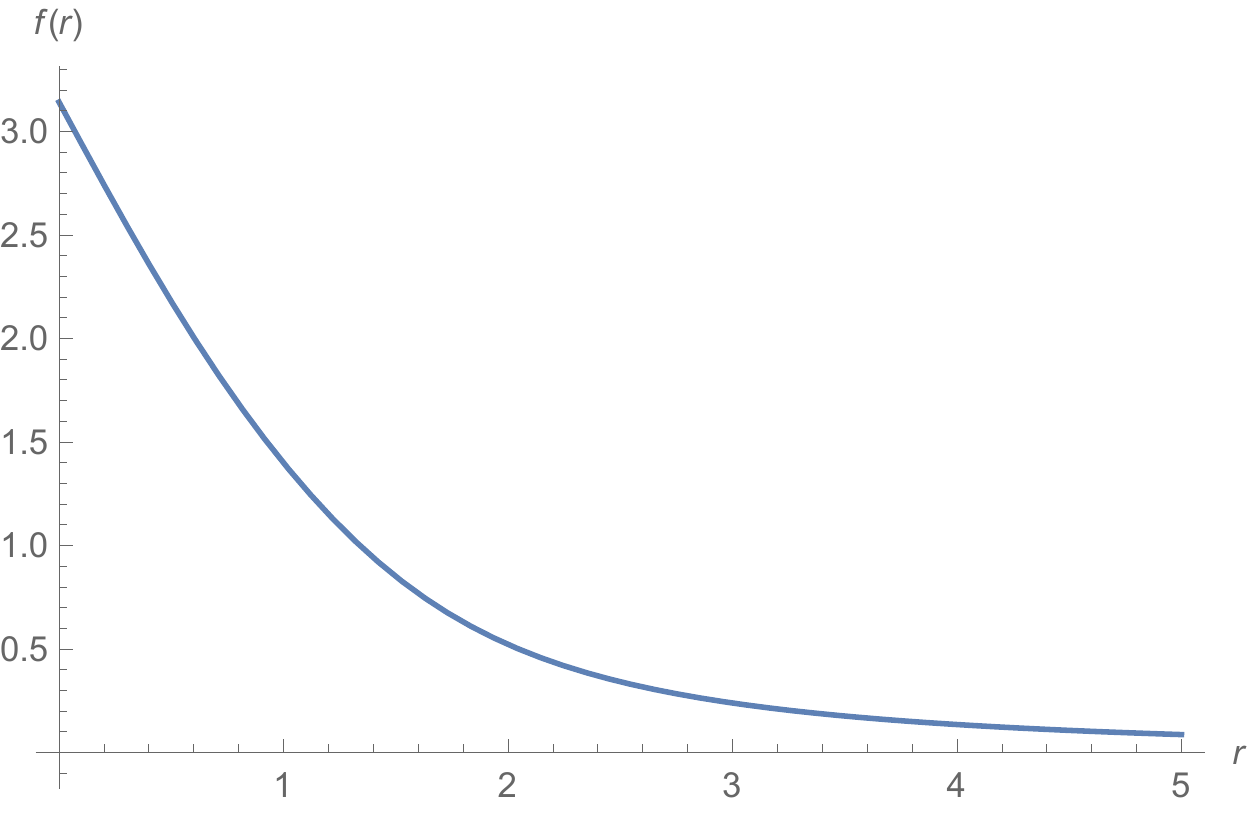}
}
\subfigure[The energy density plot.]
{
\includegraphics[scale=.6]{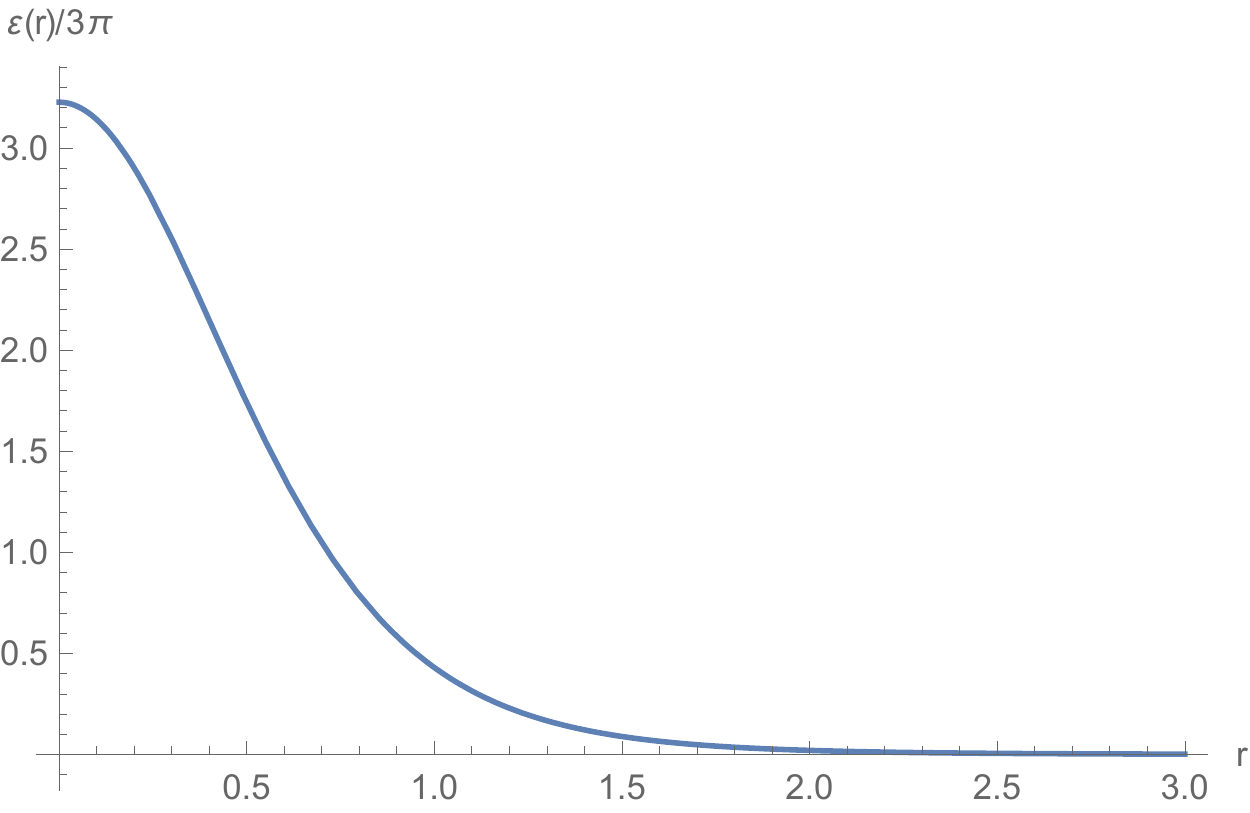}
}
\end{center}
\caption{
The numerical profile for $f(r)$ and the plot for the energy density
 $\mathcal{E} (r)$.
}
\label{fig:numerical4d}
\end{figure}

Now it is easy to show that the Sutcliffe's truncation of the Yang-Mills
action \eqref{eq:4dYM} gives the energy functional for the static Skyrme field.
Plugging the decomposition \eqref{eq:gauge_Skyrme} into the quadratic
Yang-Mills action \eqref{eq:4dYM} and performing the integration over $x^4$, then we find
\begin{align}
S =  \frac{1}{\kappa g^2} \int \! d^3 x \
\left(
- \frac{c_1}{2}
\text{Tr}
[
R_i R_i
]
- \frac{c_2}{16}
\text{Tr}
[
R_i,R_j
]^2
\right),
\label{eq:3dSutcliffe_energy}
\end{align}
where the numerical factors are calculated as $c_1 = \frac{1}{4\sqrt{\pi}}\simeq0.141$, $c_2 = 
2\int_{-\infty}^{\infty} \! dx^4 \psi_{+}^2 (\psi_+ - 1)^2 \simeq 0.198$.
These numerical factors can be set to $c_1 = c_2 = 1$ by the
rescalings of the length $x^i \to \sqrt{c_2/c_1}x^i$ and the overall factor of the action $S \to
\frac{1}{\sqrt{c_1 c_2}} S$. We therefore consider the natural unit $c_1 = c_2 = 1$ and
set $\kappa = 1, g = 1$ for simplicity.
After the rescaling, the action \eqref{eq:3dSutcliffe_energy} becomes
the energy functional for the static Skyrme field:
\begin{align}
E_{\text{Skyrme}} = \int \! d^3 x \
\left(
-\frac{1}{2} \text{Tr} [R_i R_i] - \frac{1}{16} \text{Tr} [R_i, R_j]^2
\right).
\label{eq:4dSkyrme_energy}
\end{align}
The Bogomol'nyi completion of the energy functional
\eqref{eq:4dSkyrme_energy} gives the energy bound
$E_{\text{Skyrme}} \ge 12 \pi^2 |B|$ where $B = - \frac{1}{24 \pi^2}
\int \! d^3 x \ \varepsilon_{ijk} \text{Tr} [R_i  R_j R_k]$ is the
topological charge, namely, the Baryon number.
Here $\varepsilon_{ijk}$ is the Levi-Civita symbol.
The equation of motion derived from \eqref{eq:4dSkyrme_energy} is
\begin{align}
\partial_i
\left(
R_i - \frac{1}{4} [R_j, [R_j, R_i]]
\right) = 0.
\label{eq:eom4d}
\end{align}
No analytic solutions to this equation have been found
but a spherically symmetric solution is dealt with the following hedgehog
ansatz:
\begin{align}
U = \exp
\left(
i f (r) \hat{x}^i \tau_i
\right).
\label{eq:hedgehog4d}
\end{align}
Here $\hat{x}^i = \frac{x^i}{r}$, $r^2 = x^i x^i$ and $\tau^i$ are the
Pauli matrices, namely, the quaternion basis.
The energy functional for this ansatz is evaluated to be
\begin{equation}
E_{\text{Skyrme}}=\int_0^{\infty}dr\int_{S^2}d\Omega_2\mathcal{E}(r) = 2\pi\int_0^{\infty}dr \left( r^2(\partial_rf)^2 + 2\sin^2f\left( 1+(\partial_rf)^2 \right) + \frac{\sin^4f}{r^2} \right).
\end{equation}
Here $\mathcal{E}(r)$ is the energy density and $d\Omega_2$ is the integral element of the two-dimensional sphere.
The boundary condition is given by $f(0) = \pi$, $f (\infty) = 0$.
The numerical study is easily performed for this ansatz.
The solution to the equation of motion \eqref{eq:eom4d} with the ansatz
\eqref{eq:hedgehog4d} is found in Fig. \ref{fig:numerical4d}.
The solution in Fig. \ref{fig:numerical4d} has the Baryon number $B = 1$.

We note that the energy functional \eqref{eq:4dSkyrme_energy} breaks the
scale invariance presented in the Yang-Mills action.
A physical origin of this violation comes from the Sutcliffe's truncation
\eqref{eq:gauge_Skyrme} where only the zero-mode (Skyrme field) is taken into
account. Once we include all the vector meson modes $W_i^n$, the
scale invariance is expected to be recovered.

\subsection{Eight-dimensional Skyrme model}
Now we generalize the procedure in the previous subsection to eight dimensions.
In eight dimensions, the natural action whose BPS equation is the
self-duality equation $F \wedge F = *_8 (F \wedge F)$ is that of the
{\it quartic Yang-Mills} theory. The action is
\begin{align}
S_{\text{quartic}} =& \ \frac{\alpha}{\kappa g^2} \int \! \text{Tr}
\left[
*_8 (F \wedge F) \wedge (F \wedge F)
\right]
\notag \\
=& \ \left(\frac{1}{2!}\right)^4 \frac{4}{4!}
\frac{\alpha}{\kappa g^2} \int \! d^8 x \ \text{Tr}
\left[
(F^{MN} F^{PQ})^2 - 4 F^{MN} F^{PQ} F_{MP} F_{NQ} + (F^{MN} F_{MN})^2
\right].
\label{eq:quartic_YM}
\end{align}
Here $M,N, \ldots = 1, \ldots, 8$
and the component of the gauge field strength 2-form $F = \frac{1}{2!} F_{MN} d x^M \wedge d
x^N$ is $F_{MN} = \partial_M A_N - \partial_N A_M + [A_M, A_N]$.
A constant $\alpha$ has mass dimension $[\alpha] = -4$ and
$g$ is the gauge coupling constant whose mass dimension is $-2$.
In the following, we set $\alpha/g^2 = 96$ and $\kappa = 1$ for simplicity.
The gauge field $A_M$ is in the adjoint representation of a Lie algebra associated with the
gauge group $G$. 
We consider a gauge group $G$ which admits a non-trivial homotopy $\pi_7
(G) = \mathbb{Z}$.
\footnote{
In order that instantons are classified by the integer topological charge,
it is necessary that the homotopy group contains at least one
$\mathbb{Z}$ factor.
For example, we can consider the gauge group $SO(8)$ in that case we have $\pi_7 (SO(8)) =
\mathbb{Z} \times \mathbb{Z}$.
}.

The analysis is completely parallel to the four-dimensional case.
We decompose the directions $x^M = (x^I, x^8), \ (I = 1, \ldots, 7)$ and
expand the gauge field in terms of the Hermite function $\psi_n (x^8)$.
The Sutcliffe's truncation provides the static Skyrme field in eight dimensions
through the relations \eqref{eq:gauge_Skyrme}.
Plugging the expansion \eqref{eq:gauge_Skyrme} into the quartic
Yang-Mills action \eqref{eq:quartic_YM} and performing the integration
over the $x^8$-direction,
we obtain the energy functional for the static Skyrme field:
\begin{align}
E_{\text{Skyrme}}
&=\int~d^7x\text{Tr}\Bigl[ c_2\left( [R_I,R_J][R_I,R_J] \right)^2 + c_2\left( [R_I,R_J][R_K,R_L] \right)^2 \notag \\
&\hspace{100pt}- 4c_2[R_I,R_J][R_K,R_L][R_I,R_K][R_J,R_L] \notag \\
&\hspace{30pt} + 4c_1\left( [R_I,R_J] \right)^2R_K^2 + 4c_1\left( [R_I,R_J]R_K \right)^2 - 4c_1[R_I,R_J]R_K[R_I,R_K]R_J \notag \\
&\hspace{100pt}+ 8c_1[R_I,R_J][R_K,R_I]R_JR_K - 4c_1[R_I,R_J]R_I[R_K,R_J]R_K \Bigr].\label{eq:8dSkyrme}
\end{align}
Here $R_I = U\partial_I U^{-1}$
is the right current and
the Skyrme field is defined by the holonomy
\begin{align}
U (x^I) = -\text{P} \exp
\left[
\int^{\infty}_{-\infty} \! d x^8 \ A_8 (x^I, x^8)
\right].
\end{align}
Therefore, the Skyrme field is a map $U : \mathbb{R}^7 \mapsto \tilde{G}$ where $\tilde{G}$ is a group manifold.
The numerical constants $c_1$, $c_2$ in \eqref{eq:8dSkyrme} are calculated to be
\begin{align}
c_1 = \int^{\infty}_{-\infty} \! dx^8 \ \frac{1}{2 \sqrt{\pi}} \psi_0^2
 \psi_+^2 (\psi_+ - 1)^2 \simeq 0.00940,
 \quad
c_2 = \int^{\infty}_{-\infty} \! dx^8 \ \psi_+^4 (\psi_+-1)^4 \simeq 0.00308.
\end{align}
As in the case of the four-dimensional Skyrme model, these numerical
factors are scaled away by the replacements $x^I \to \sqrt{c_2/c_1}x^I$,
$E_{\text{Skyrme}} \to \frac{1}{\sqrt{c_1 c_2}} E_{\text{Skyrme}}$.
We therefore set $c_1 = c_2 = 1$.
The quartic Yang-Mills action \eqref{eq:quartic_YM} has the scale
invariance while the energy functional \eqref{eq:8dSkyrme} does not.
Again, this is due the Sutcliffe's truncation where only the
zero-mode is considered and the vector mesons are neglected.

The eight-dimensional Skyrme model \eqref{eq:8dSkyrme} has similar
properties with the four-dimensional ones.
For example, the energy functional \eqref{eq:8dSkyrme} is invariant under the
following global transformation
\begin{align}
U \to O_L U O_R^{-1}, \qquad O_L, O_R \in \tilde{G}.
\end{align}
This is a generalization of the chiral symmetry in four dimensions.
One also finds that the energy functional \eqref{eq:8dSkyrme} consists of the
terms with 6th and 8th derivatives.
This is compared with the 2nd and 4th derivative terms in the
four-dimensional Skyrme model.
The Derrick's theorem applied to the energy \eqref{eq:8dSkyrme} indicates
that there is a stable solitonic solution to this model.
We call this the eight-dimensional Skyrmions.
The Bogomol'nyi completion of the energy \eqref{eq:8dSkyrme} is given by
\begin{align}
E_{\text{Skyrme}} =& \ 4\int \! d^7 x \ \text{Tr}
\Bigg[
\left(
\sqrt{\frac{1}{3!}} \varepsilon_{IJKLABC} R_I R_J R_K \pm \sqrt{4!} R_{[L} R_A R_B R_{C]}
\right)^2
\notag \\
& \qquad \qquad \qquad
\mp 4 \varepsilon_{IJKLABC} R_I R_J R_K R_L R_A R_B R_C
\Bigg] \ge \frac{16}{N_C} |\mathcal{B}|,	\label{eq:8dSkyrme_BPS_bound}
\end{align}
where $N_C = 1/9600\pi^{4}$ is the normalization constant of the following topological charge:
\begin{align}
\mathcal{B} = N_C \int \! d^7 x \ \text{Tr}
\left[
\varepsilon_{IJKLABC} R_I R_J R_K R_L R_A R_B R_C
\right].	\label{eq:8dSkyrme_topological_charge}
\end{align}
Here $\varepsilon_{IJKLABC}$ is the totally antisymmetric tensor.
The topological charge \eqref{eq:8dSkyrme_topological_charge}
is the natural generalization of the Baryon number $B = -
\frac{1}{24 \pi^2} \int \! d^3 x \ \text{Tr} [\varepsilon_{ijk} R_i R_j R_k]$
in the four-dimensional Skyrme model.

\section{Eight-dimensional Skyrmions from instantons}
In this section, we examine a field configuration that extremizes the energy
functional \eqref{eq:8dSkyrme}, namely, the Skyrmion in eight
dimensions.
Assuming the hedgehog ansatz for the Skyrme field $U(x)$, we first
derive the equation of motion from \eqref{eq:8dSkyrme}.
We will find a solution to the equation by the numerical analysis.
We then construct a field configuration from the eight-dimensional
instantons through the Atiyah-Manton prescription.
We compare the two solutions and verify whether the Atiyah-Manton
approximation works even in eight dimensions.

\paragraph{Skyrmions from numerical analysis}
Following the standard scheme for a spherically symmetric
solution to the four-dimensional Skyrme model, we consider the following
hedgehog ansatz:
\begin{align}
U (x) = \exp
\left(
f (r) \hat{x}^I e^{\dagger}_I
\right),
\label{eq:hedgehog}
\end{align}
where $\hat{x}^I = \frac{x^I}{r}$, $r^2 = x^I x^I$
and $f(r)$ is a real function.
The basis $e_I, e_I^{\dagger}$ is the higher dimensional analogue of
the pure imaginary quaternions in four dimensions.
Note that we do not employ the octonions as a higher
dimensional generalisation of the quaternions.
It is well known that the octonions are never represented by matrices and
the algebra based on them loses the associativity \cite{Baez:2001dm}.
The natural candidate for the basis in eight dimensions is based on the
Clifford algebra \cite{Nakamula:2016srw}.
This is given by
\begin{align}
e_{M} = \delta_{M 8} \mathbf{1}_8 + \delta_{MI} \Gamma_I^{(-)}, \qquad
e_{M}^{\dagger} = \delta_{M 8} \mathbf{1}_8 + \delta_{MI} \Gamma_I^{(+)},
\qquad
(M = 1, \ldots, 8, \ I= 1, \ldots, 7),
\label{eq:8d_basis}
\end{align}
where $\Gamma_I^{(\pm)}$ are $8 \times 8$ matrices that satisfy the
relations $\{ \Gamma_I^{(\pm)}, \Gamma_J^{(\pm)} \} = -2 \delta_{IJ} \mathbf{1}_8$.
The matrices $\Gamma_I^{(\pm)}$ are defined by $\Gamma_I^{(\pm)} = \frac{1}{2}(1\pm\omega)\Gamma_I$.
We choose the matrices $\Gamma^{(\pm)}_I$ such that they satisfy the relation $\Gamma_I^{(+)}=-\Gamma_I^{(-)}$.
Here $\Gamma_I$ are given by the matrix representation of the
seven-dimensional complex Clifford algebra $\Gamma_I \in C\ell_7
(\mathbb{C})$ and $\omega = (-1) \Gamma_1 \cdots
\Gamma_7$ is a chirality matrix.
The basis is normalized as $\text{Tr} [e_M e^{\dagger}_N] = 8
\delta_{MN}$ and satisfies the following relations
\begin{align}
& e_M e^{\dagger}_N + e_N e^{\dagger}_M = e_M^{\dagger} e_N +
 e_N^{\dagger} e_M = 2 \delta_{MN} \mathbf{1}_8,
\notag \\
& e_M e_N + e_N e_M = 2 \delta_{M8} e_N + 2 \delta_{N8} e_M - 2
 \delta_{MN} \mathbf{1}_8,
\notag \\
& e^{\dagger}_M e^{\dagger}_N + e^{\dagger}_N e^{\dagger}_M = 2
 \delta_{M8} e^{\dagger}_N + 2 \delta_{N8} e^{\dagger}_M - 2 \delta_{MN}
 \mathbf{1}_8.
\label{eq:algebra8d}
\end{align}
Note that we have $e^{\dagger}_I = - e_I$ in our construction.
Therefore the hedgehog field configuration \eqref{eq:hedgehog} satisfies $U^{\dagger} U =
\mathbf{1}_8$ and it belongs to $U(8)$.
The details of the Clifford algebra, including the explicit matrix representations of the basis $e_M,
e^\dagger_M$, are found in \cite{Nakamula:2016srw}.

We now derive the equation of motion for the profile function $f (r)$.
Using the algebra of the basis \eqref{eq:algebra8d}, we find that the hedgehog ansatz is expanded
as
\begin{align}
U(x) = \cos f \mathbf{1}_8 + \sin f \hat{x}^I e_I^{\dagger}.
\end{align}
This expression allows us to write down the right-current field:
\begin{align}
R_I = - r^{-1} \sin^2 f \hat{x}_I \mathbf{1}_8 +
(-r^{-1} \sin f \cos f + \partial_r f) \hat{x}_I \hat{x}^{\dagger}
+r^{-1} \sin f \cos f e^{\dagger}_I - r^{-1} \sin^2 f e^{\dagger}_I \hat{x}^{\dagger}.
\end{align}
Here $\hat{x} = \hat{x}^I e_I$, $\hat{x}^{\dagger} = \hat{x}^I
e_I^{\dagger}$.
It is straightforward to calculate each term in \eqref{eq:8dSkyrme}
by using the above expression and the algebra associated with the basis \eqref{eq:algebra8d}.
Although the derivation is tedious, it needs a little bit of effort.
The details are found in appendix.
The energy functional becomes, 
\begin{align}
E_{\text{Skyrme}} &=
 \int_0^{\infty}dr\int_{S^6}d\Omega_6\mathcal{E}(r) \notag \\
 &= 24576\pi^3 \int^{\infty}_0 \! dr \
\left(
3  r^2 \sin^4 f (\partial_r f)^2 + 4 \sin^6 f (4 (\partial_r f)^2
 + 1) + 12 \frac{\sin^8 f}{r^2}
\right),	\label{eq:8dSkymre_action_hedgehog}
\end{align}
where the overall factor comes from the volume factor of the radial direction and algebras containing $e_M,e^{\dagger}_M$ (see appendix).
Then, we derive the equation of motion for $f(r)$ as
\begin{align}
& \sin^2 f (3 r^2 + 16 \sin^2 f) \partial_r^2 f + 6 r \sin^2 f
 \partial_r f
\notag \\
& \qquad
+ 3 \sin 2f
\left[
(r^2 + 8 \sin^2 f) (\partial_r f)^2 - 2 \sin^2 f - 8 \frac{\sin^4 f}{r^2}
\right] = 0.
\label{eq:eom}
\end{align}
The boundary condition for the profile function $f (r)$ is
\begin{align}
f (0) = \pi, \qquad f (\infty) = 0.
\label{eq:bc}
\end{align}

Compared with the equation in four dimensions, the equation \eqref{eq:eom}
looks highly non-linear. Therefore it is not obvious whether the equation
\eqref{eq:eom} has appropriate solutions that are consistent with the boundary condition
\eqref{eq:bc} or not.
In order to clarify the existence of the solution to the equation
\eqref{eq:eom}, we first perform the Taylor expansion of the profile function at
the origin: $f(\delta r) = \sum_{i=0}^{\infty}f_i(\delta r)^i = f_0 +
f_1\delta r + f_2(\delta r)^2 + \dots$.
We then write down the equations for the coefficients $f_i$ and look for
$f_i$ order by order in $(\delta r)^i$.
For the boundary condition \eqref{eq:bc}, we find that the asymptotic
behavior of the solution at the origin is
\begin{align}
f(\delta r) = \pi + f_1\delta r - \frac{\left( 3c_1+8c_2f_1^2
 \right)f_1^3}{9\left( 3c_1+16c_2f_1^2 \right)} (\delta r)^3 +
 \mathcal{O}\left( (\delta r)^5 \right).
\label{eq:8d_Skyrme_EOM_expansion_origin}
\end{align}
\begin{figure}[tb]
\begin{center}
\subfigure[The profile for the 8d Skyrmion.]
{
\includegraphics[scale=.6]{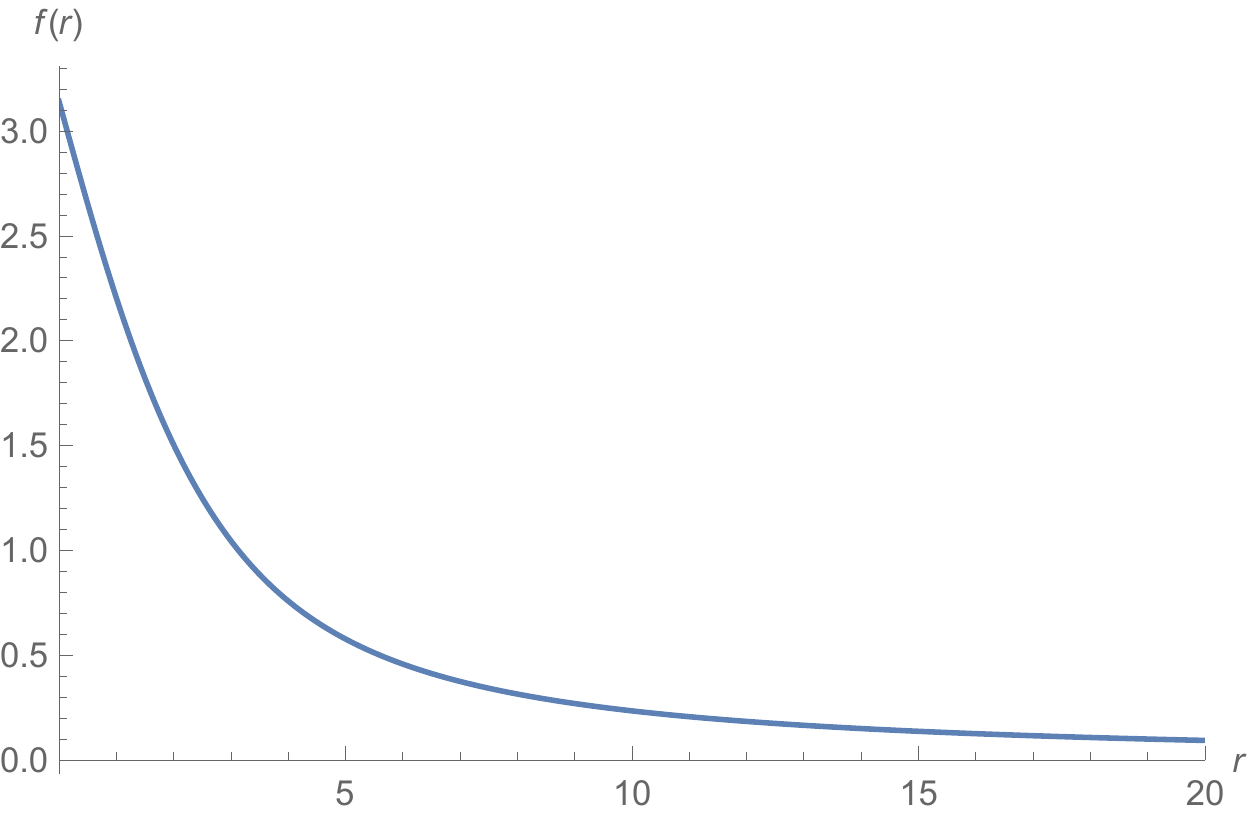}
}
\subfigure[The energy density plot for the 8d Skyrmion.]
{
\includegraphics[scale=.6]{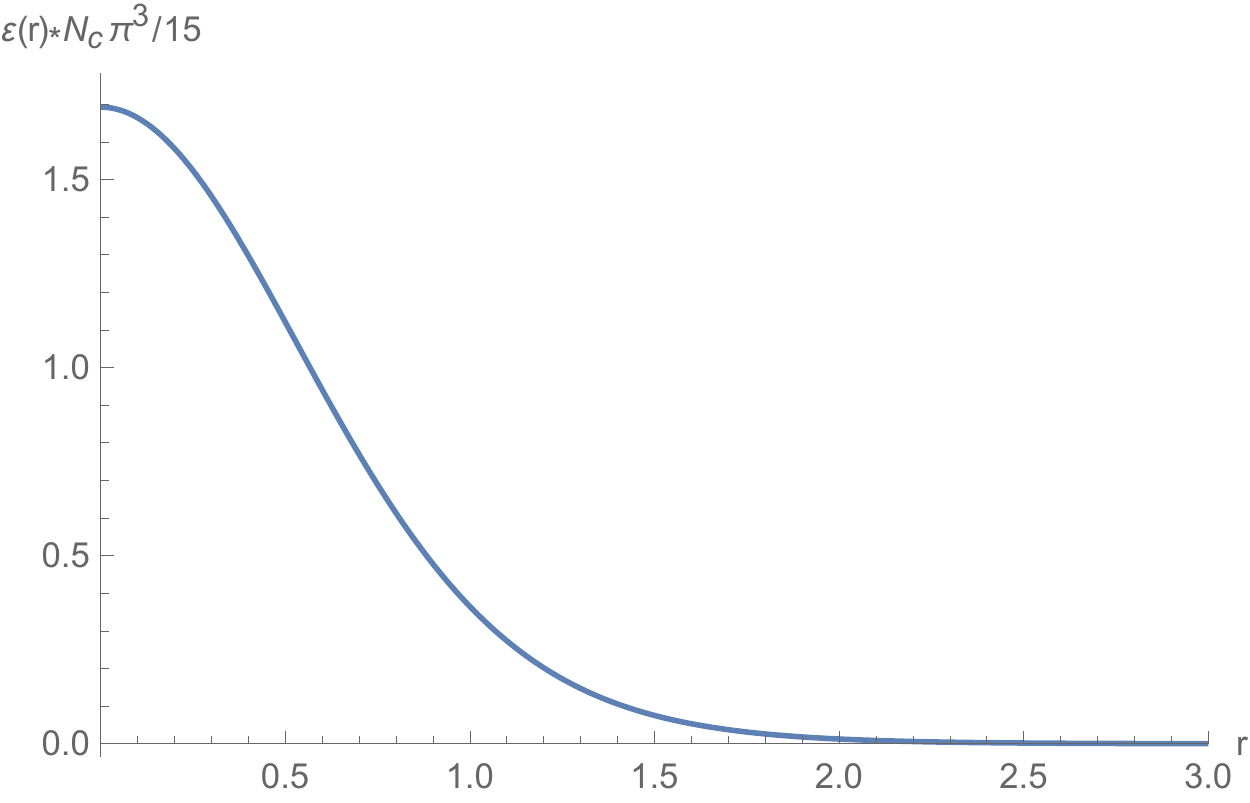}
}
\end{center}
\caption{
The numerical profile for $f(r)$ and the plot for the energy density
 $\mathcal{E} (r)$.
}
\label{fig:numerical}
\end{figure}
Here $f_1$ can be chosen as a shooting parameter in the numerical analysis.
From this observation, we conclude that we can numerically calculate a
solution to the equation \eqref{eq:eom} by appropriate methods of 
second ordinary differential equations with boundary conditions.
We stress that, if any shooting parameters are not found, then the equation does
not have appropriate solution with the boundary condition in general.
The numerical result is found in Fig. \ref{fig:numerical} where we have employed
the functional Newton-Raphson method.
The behaviour of the profile function and the energy functional is quite
similar to those in the four-dimensional Skyrmion (see Fig. \ref{fig:numerical4d}).

The Skyrme field is a map $\mathbb{R}^7 \mapsto U(8)$.
However, the boundary condition $U(r) \to \mathbf{1}_8 \ (r \to \infty)$
implies that the base manifold is topologically $S^7$.
Therefore the solutions are characterized by the topological charge associated with
the homotopy group $\pi_7 (U(8)) = \mathbb{Z}$.
Indeed, the topological charge for the hedgehog ansatz \eqref{eq:hedgehog}
and the boundary condition \eqref{eq:bc} is
evaluated to be
\begin{align}
\mathcal{B} = -9600 \pi^3 N_C ( f(\infty) - f (0) ) = 1.
\end{align}
This is the single Skyrmion in eight dimensions.

\paragraph{Atiyah-Manton solution from instantons}
We next make contact with the Skyrmion from the eight-dimensional instantons.
The Bogomol'nyi completion of the quartic Yang-Mills action
\eqref{eq:quartic_YM} is
\begin{align}
S_{\text{quartic}}
=& \ \frac{\alpha}{2\kappa g^2}
\int \! \mathrm{Tr}
\left[
(F \wedge F \mp *_8 (F \wedge F))^2 \pm 2F \wedge F \wedge F \wedge F
\right]
\notag \\
\ge& \ \pm \frac{\alpha}{\kappa g^2}
\int \! \mathrm{Tr} [F \wedge F \wedge F \wedge F].
\label{eq:quarticBPS}
\end{align}
Here we have defined
\begin{align}
(F \wedge F \pm *_8 F \wedge F)^2
=
(F \wedge F \pm *_8 F \wedge F) \wedge *_8
(F \wedge F \pm *_8 F \wedge F).
\end{align}
The action is bounded from below by the fourth Chern number $k = \int
\text{Tr} [ F \wedge F \wedge F \wedge F]$ which defines the topological
charge associated with instantons.
The theory defined by the action \eqref{eq:quarticBPS} has scale invariance.
The Derrick's theorem implies that the theory admits static
solitons, namely, instantons.
The Bogomol'nyi bound is saturated when the (anti-)self-duality equation
\begin{align}
F \wedge F = \pm *_8 F \wedge F,
\label{eq:SD}
\end{align}
is satisfied. 
This is a natural generalization of the (anti-)self-duality equation $F
= \pm *_4F$ in four dimensions.
In the following we choose the plus sign in \eqref{eq:SD}.
Solutions to the equation \eqref{eq:SD} is known as the self-dual instantons in eight
dimensions. They are characterized by the homotopy group $\pi_7 (G) =
\mathbb{Z}$ where $G$ is a gauge group.
Only the one-instanton is known as an analytic solution in the past
\cite{Grossman:1984pi,Tchrakian:1984gq}
\footnote{
We note that multi-instanton solutions to the self-duality equation \eqref{eq:SD} are discussed in
the framework of the ADHM construction \cite{Nakamula:2016srw}.
}.
The one-instanton solution is given by
\begin{align}
A_{M}
= \frac{1}{4}\partial_N\ln\left(1+\frac{\lambda^2}{\|\tilde{x}\|^2}\right)\Sigma_{MN}^{(-)},
\label{eq:1instanton}
\end{align}
where $\|\tilde{x}\|^2 = (x^M-a^M)(x_M-a_M)$ and
$\lambda$, $a^M$ are the size and the position moduli of the solution.
For simplify, we set $a^M=0$.
The matrix
\begin{align}
\Sigma^{(-)}_{MN} = e_{M} e_{N}^{\dagger} - e_{N} e^{\dagger}_{M}
\end{align}
is the generator of the $SO(8)$ Lorentz group.
This is the eight-dimensional analogue of the
't Hooft instanton in four dimensions \cite{'tHooft:1976fv}.

Following Atiyah and Manton \cite{Atiyah:1989dq}, we calculate the
holonomy for the instanton solution \eqref{eq:1instanton}.
To this end, it is convenient to rewrite the solution
\eqref{eq:1instanton} as
\begin{align}
A_{M} (x^I,x^8)=& \ \frac{1}{2}
\left(
\frac{1}{\lambda^2 + r^2 + (x^8)^2} - \frac{1}{r^2 + (x^8)^2}
\right) x^{N} \Sigma_{MN}^{(-)}.
\end{align}
Then one finds
\begin{align}
A_8 =
\left(
\frac{1}{\lambda^2 + r^2 + (x^8)^2} - \frac{1}{r^2 + (x^8)^2}
\right) x^I e_I^{\dagger}.
\end{align}
\begin{figure}[tb]
\centering
\includegraphics[scale=0.8]{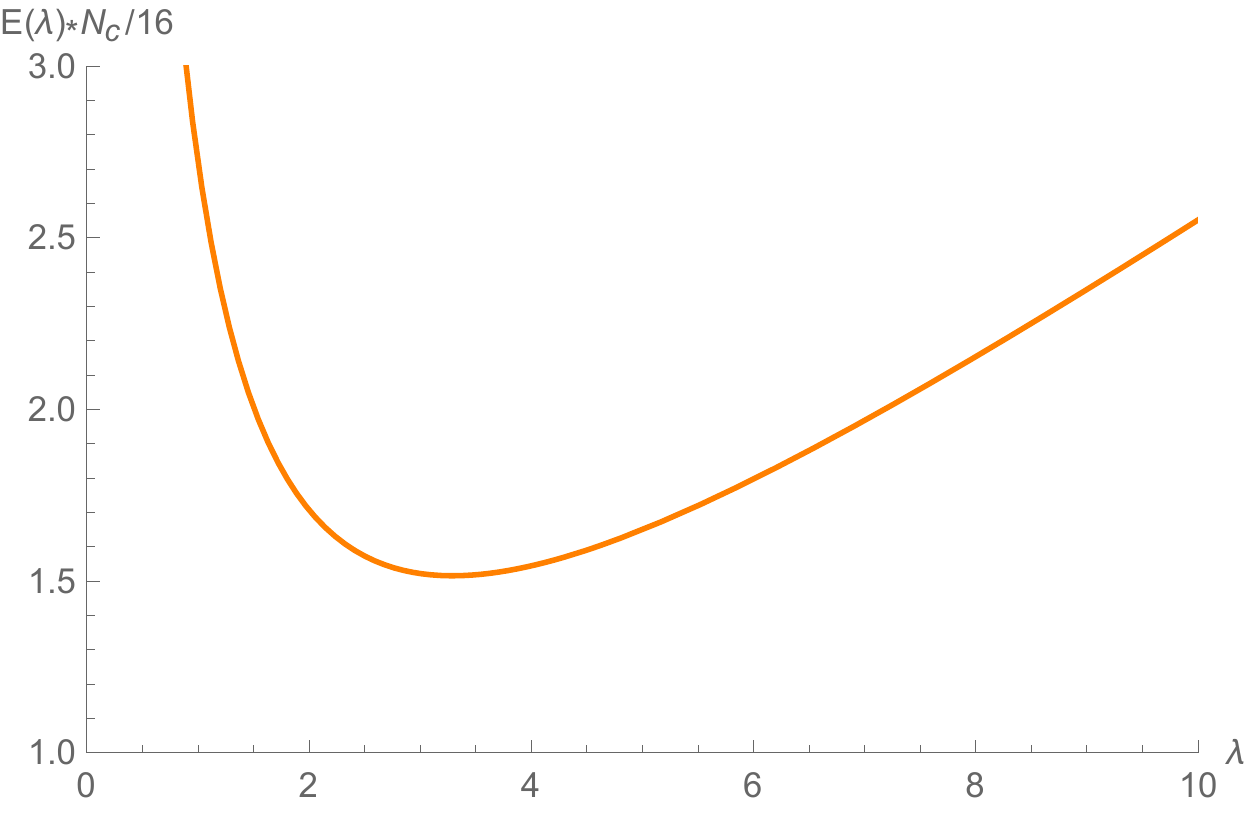}
\caption{
The energy profile for the Atiyah-Manton solution \eqref{eq:instanton_skyrmion}
as the function of the instanton size $\lambda$.
}
\label{fig:AM_energy_size}
\end{figure}
Using this representation, we calculate the following holonomy for the
one-instanton solution $A_8$:
\begin{align}
U (x^I) = -\text{P} \exp \int^{\infty}_{- \infty} \! d x^8 \ A_8 (x^I,x^8)
= \exp
\left[
\pi
\left(
1-\frac{r}{ \sqrt{r^2+\lambda^2} }
\right) \hat{x}^I e_I^{\dagger}
\right].
\label{eq:instanton_skyrmion}
\end{align}
The result is the standard hedgehog form for the Skyrme field \eqref{eq:hedgehog}.
This is why we have employed the basis $e_I^{\dagger}$ in \eqref{eq:hedgehog}.
Plugging the Atiyah-Manton solution \eqref{eq:instanton_skyrmion} into
the quartic Yang-Mills action \eqref{eq:quartic_YM} results in the
static energy $E(\lambda)$ for the solution.
The plot for $E(\lambda)$ is found in Fig. \ref{fig:AM_energy_size}.
As anticipated, the energy depends on the size of the instanton $\lambda$.
This is because the Sutcliffe's truncation breaks the scale invariance in the quartic
Yang-Mills model. The size $\lambda$ now lost its status of modulus.
The true solution corresponds to the extremum of $E(\lambda)$.
We find this happens at $\lambda = 3.29095$.

\begin{figure}[tb]
\centering
\includegraphics[scale=0.8]{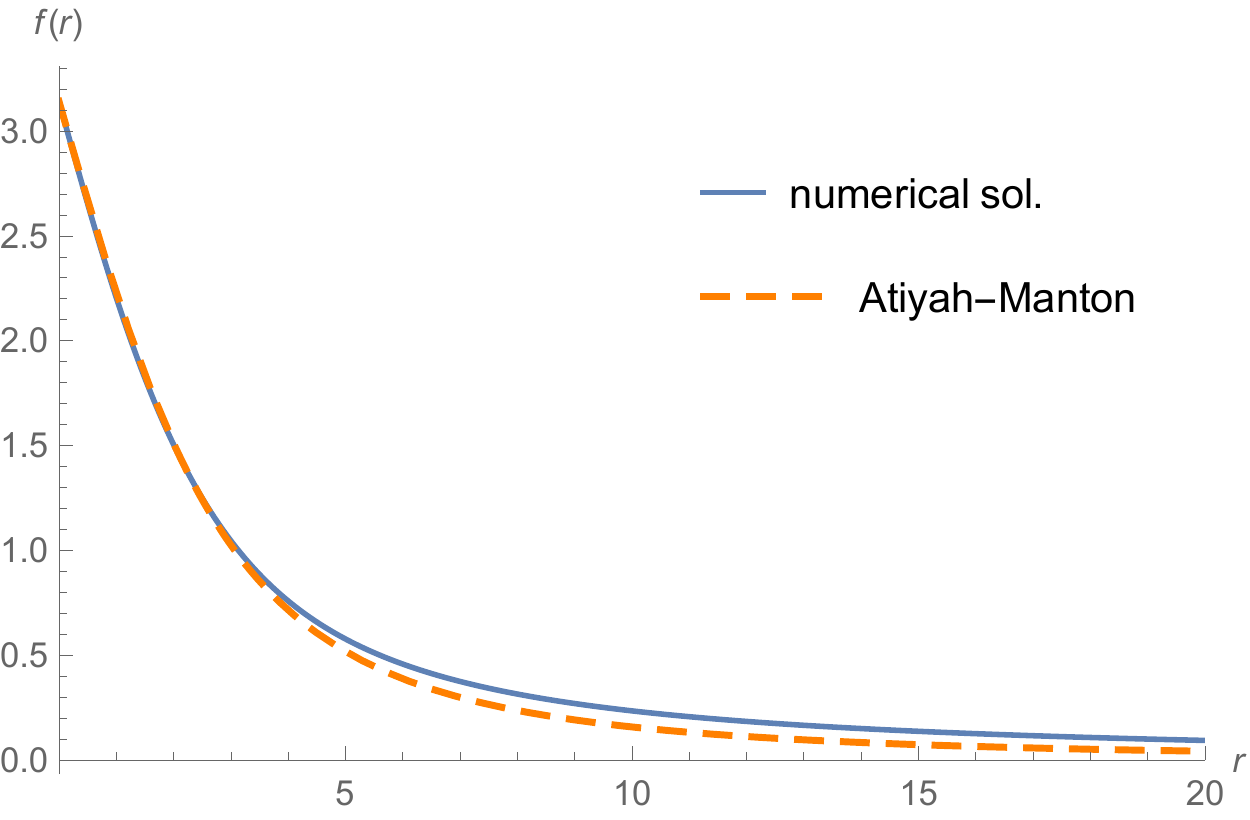}
\caption{
The profile functions for the numerical and the Atiyah-Manton solutions.
}
\label{fig:profiles}
\end{figure}

\begin{figure}[tb]
\centering
\includegraphics[scale=0.8]{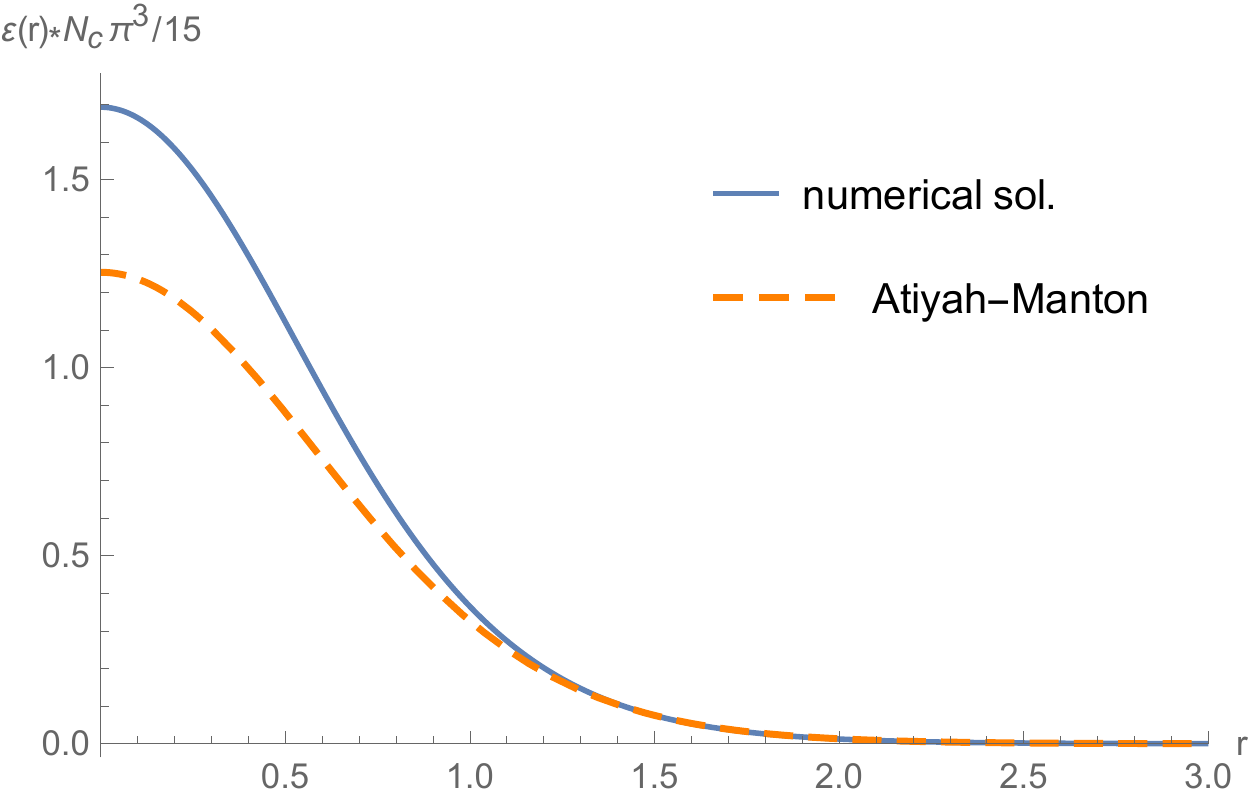}
\caption{
The profile functions for the energy density.
The numerical versus the Atiyah-Manton solutions.
}
\label{fig:energy_density}
\end{figure}

For this value of $\lambda$, we now compare the profile functions of the
Atiyah-Manton and the numerical solutions. The result is found in Fig \ref{fig:profiles}.
We find that they agrees with high accuracy.
The plot for the energy density is also compared in Fig \ref{fig:energy_density}.
Again, we find a good agreement between them.
This result can be confirmed by evaluating the total energy
(see Table \ref{tb:energy}).
We therefore conclude that the Atiyah-Manton construction of Skyrmions from instantons works well even
in eight dimensions.
We note that the eight-dimensional Skyrmion is a non-BPS solution which
is same as the four-dimensional one.
\pagebreak
\begin{table}[tb]
\centering
\begin{tabular}{|c||c|c|c|}
\hline
Solution & Numerical & Atiyah-Manton & BPS bound\\
\hline
Energy & $1.51239\times16/N_c$ & $1.51521\times16/N_c$ & $16/N_c$
\\
\hline
  \end{tabular}
\caption{The total energy for the numerical, the Atiyah-Manton
 solutions and the BPS bound in this model \eqref{eq:8dSkyrme_BPS_bound}.}
\label{tb:energy}
\end{table}

\section{Higher dimensional generalization}
In this section we make an analysis on the Atiyah-Manton construction in
$4k$ dimensions.
It is worthwhile first to mention about the $k = 3$, namely, the twelve-dimensional
case. In twelve dimensions, the self-duality equation becomes
$F \wedge F \wedge F = \pm *_{12} F \wedge F \wedge F$.
It is an easy exercise to show that the one-instanton solution to this equation is given
by \eqref{eq:1instanton} where the $SO(8)$
generator $\Sigma_{\mu \nu}^{(-)}$ is replaced by that of $SO(12)$.
We can construct the Atiyah-Manton solution by
calculating the holonomy associated with the instanton solution.
We can also find the Skyrme model in twelve dimensions and its Skyrmion
solution along the lines of the eight-dimensional case.
The discussion is parallel to that in eight dimensions presented in this paper.
However, the explicit calculation of the Sutcliffe's truncation in
twelve dimensions results in the energy functional for the Skyrme model
with diverse (about $\mathcal{O} (10^2)$) terms.
Analyzing all the terms is beyond the scope of this paper. 
We therefore proceed to the general discussion in the following.

Now we move to the discussion in $4k$ dimensions.
The $4k$-dimensional generalization of the quartic Yang-Mills action
\eqref{eq:quartic_YM} is
\begin{align}
S_{\text{YM}} = \int_{\mathbb{R}^{4k}} \! \text{Tr} [
 (*_{4k} F^k) \wedge (F^k) ],
\end{align}
where $F^k$ is the $k$th wedge products of the gauge field strength
2-form, $F^k = F \wedge \cdots \wedge F$.
The gauge field takes value in the adjoint representation of a gauge
group $G$.
We assume that this gauge group has non-trivial homotopy $\pi_{4k-1} (G)
= \mathbb{Z}$.
It is straightforward to perform the Bogomol'nyi completion of the
action:
\begin{align}
S_{\text{YM}} = \frac{1}{2} \int_{\mathbb{R}^{4k}} \! \text{Tr}
[
(F^k \mp *_{4k} F^k)^2 \pm 2F^{2k}
] \ge
\pm \int_{\mathbb{R}^{4k}} \! \text{Tr}
[
F^{2k}
].
\end{align}
The BPS equation becomes
\begin{align}
F^{k} = \pm *_{4k} F^k.
\end{align}
This is the (anti-)self-duality equation in $4k$ dimensions.
The one-instanton solution to this equation is explicitly wrote down by the
ADHM construction of instantons in $4k$ dimensions \cite{Takesue} which
is the $4k$-dimensional generalization of \cite{Nakamula:2016srw} in
eight dimensions.
Again, the solutions are given as the form in \eqref{eq:1instanton}
where the $SO(8)$ generator is replaced by those of $SO(4k)$.

Next we perform the Sutcliffe's truncation.
The index structure of the Yang-Mills Lagrangian is
\begin{align}
& (*_{4k} F^k) \wedge F^k \notag \\
& = \frac{1}{(2k)!}
\left(
\frac{1}{2!}
\right)^{2k}
\varepsilon_{M_1 \cdots M_{2k} N_1 \cdots N_{2k}} \varepsilon^{M_1
 \cdots M_{2k} P_1 \cdots P_{2k}} F^{N_1 N_2} \cdots F^{N_{2k-1} N_{2k}}
 F_{P_1 P_2} \cdots F_{P_{2k-1} P_{2k}} d^{4k} x,
\end{align}
where the overall factor comes from the normalization of the 2-form $F =
\frac{1}{2!} F_{MN} dx^M \wedge dx^N$ and the definition of the Hodge
dual operation.
The procedure of the reduction is parallel to the previous sections.
We can reduce the gauge field along, say, the $x^{4k}$-direction.
Then, the gauge field becomes
\begin{align}
& F_{I \sharp} = R_I \frac{\psi_0 (x^{4k})}{\sqrt{2} \pi^{\frac{1}{4}}},
\quad
F_{IJ} = [R_I, R_J] \psi_+ (x^{4k}) (\psi_+ (x^{4k}) - 1),
\notag \\
& (I,J, \ldots = 1, \ldots 4k-1, \ \sharp = 4k).
\end{align}
Here $R_I = U\partial_I U^{\dagger}$ is the right current field constructed
from the Skyrme field $U(x^I)$.
Then, the energy functional for the static field $U(x^I)$ in $4k$ dimensions has the following
structure
\begin{align}
E_{\text{Skyrme}} = E_{4k} (x) + E_{4k-2} (x),
\label{eq:4k_energy}
\end{align}
where $E_n$ stands for terms that contain $n$-th derivatives.
The energy \eqref{eq:4k_energy} is compared with that in the
eight-dimensional Skyrme model.
Again, the Derrick's theorem implies that there is a static soliton
solution that extremizes the energy \eqref{eq:4k_energy}.
This is nothing but the Skyrmion in $4k$ dimensions.
Finding the explicit solutions need the numerical analysis in each
dimension.
We can also calculate the holonomy for the 1-instanton solution in $4k$
dimensions and derive the static energy $E (\lambda)$.
Although we do not repeat the same calculations, the result of the original
Atiyah-Manton construction in four dimensions and our result
in eight dimensions strongly suggest that this instanton/Skyrmion correspondence does hold in
$4k$ dimensions.

\section{Conclusion and discussions}
In this paper we studied the Atiyah-Manton construction of Skyrmions in
eight dimensions. Following the formalism developed in
\cite{Sutcliffe:2010et}, we derive the static energy functional for the
Skyrme field from the quartic Yang-Mills action in eight dimensions.
The Derrick's theorem indicates that there exist stable soliton
solutions.
The solutions are classified by the topological charge $\mathcal{B} = N_C \int \! d^7
x \ \varepsilon_{IJKLABC} \text{Tr} [R_I R_J R_K R_L R_A R_B R_C]$ which
is the eight-dimensional generalization of the Baryon number.
Assuming the spherically symmetric hedgehog ansatz, we derive
the equation of motion for the Skyrme field.
Although the equation is quite nonlinear and complicated, we can perform
the numerical analysis to find solutions.
We present the explicit numerical solution for the Skyrmion associated with the
topological charge $\mathcal{B}=1$.
The profile function and the energy density of the eight-dimensional
Skyrmion look quite similar to those in four dimensions.

In the latter part of the paper, we study the relation between the
eight-dimensional instantons and the Skyrmions.
This is a generalization of the Atiyah-Manton construction in
four-dimensions. Following the four-dimensional case, we constructed the
Atiyah-Manton solution for the Skyrmion from the one instanton
solution found in \cite{Grossman:1984pi,Tchrakian:1984gq}.
We then compare the numerical solution and the Atiyah-Manton solution
and find that there is a good agreement between them.
The profile function looks quite similar in these solutions.
This result dictates us that the correspondence between the instantons
and the Skyrmions by the Atiyah-Manton construction is an universal
property in higher dimensions.

Indeed, we have confirmed that the Sutcliffe's truncation of the higher
dimensional generalization of the quartic Yang-Mills action gives the
energy functional $E$ for the static Skyrme field in $4k$ dimensions.
The structure of $E$ together with the Derrick's theorem implies that
there are stable Skyrmion solutions in $4k$ dimensions.
Since it is easy to show that the one-instantons in $4k$ dimensions are
given by the 't Hooft type, we can easily write down the Atiyah-Manton
solution in each dimension. Although it is hard to compute the explicit
energy functional for the $4k$-dimensional Skyrme model, we expect the
Atiyah-Manton solution provides a good approximation to the Skyrmions.

Our study exhibits a deep relation between Yang-Mills instantons and
Skyrmions.
Physical interpretations of the Atiyah-Manton construction in lower
dimensions are studied intensively \cite{Kudryavtsev:1997nw,
Nitta:2012xq, Nitta:2012wi, Eto:2015uqa}.
Analogous relations among various solitons are expected in higher dimensions.
Meanwhile, supersymmetry play an important role to study the BPS nature of
classical solutions. Among other things, supersymmetric generalizations
of Skyrme model in four dimensions have been studied
\cite{Bergshoeff:1984wb,Freyhult:2003zb,Adam:2013awa,Queiruga:2015xka,Gudnason:2015ryh,Gudnason:2016iex}.
It is interesting to supersymmetrize the eight dimensional Skyrme model
presented in this paper.

There are various related studies.
It is known that Skyrmions and monopoles have similar structures through
the rational map ansatz. One can expect that this relation holds even in
higher dimensions.
For example, we know that only the numerical solutions of monopoles in
seven dimensions \cite{Radu:2005rf}.
It is interesting to study the Nahm construction of monopoles
\cite{Nahm:1979yw} to find analytic solutions in seven dimensions.
These expectations may be based on the integrable structure of the
self-duality equations.
It is known that the self-duality equation in four dimensions are reduced to integrable
equations in lower dimensions \cite{Mason:1991rf}.
It is also interesting to study the integrable structure of the
self-duality equations in $4k$ dimensions and generalization of the
Ward's conjecture \cite{Ward:1985gz}.
We will come back to these issues in future studies.

\subsection*{Acknowledgments}
We would like to thank Muneto Nitta and Nobuyuki Sawado for useful discussions and comments.
The work of S.~S. is supported in part by Kitasato University Research Grant for Young
Researchers.

\begin{appendix}
\section*{Derivation of the eight-dimensional Skyrme model with the hedgehog ansatz}
For later convenience we reproduced the right current as
\begin{align}
R_I &= -r^{-1}\sin^2f \hat{x}_I\mathbf{1}_8 + \left( -r^{-1}\sin f\cos f + \partial_rf \right) \hat{x}_I\hat{x}^{\dagger} + r^{-1}\sin f\cos fe_I^{\dagger} - r^{-1}\sin^2fe_I^{\dagger}\hat{x}^{\dagger}.
\end{align}
Then we find
\begin{align}
R_IR_I &= -\left( (\partial_rf)^2+6r^{-2}\sin^2f \right)\mathbf{1}_8.
\end{align}

The commutator of the current $R_I$ is calculated to be
\begin{align}
[R_I,R_J] &= -r^{-2}\sin^2f\Sigma_{IJ}^{(-)} \notag \\
&\hspace{20pt}+ 2\left( r^{-2}\sin^2f - r^{-1}\sin f\cos f\partial_rf \right)(\hat{x}_Ie_J^{\dagger}-\hat{x}_Je_I^{\dagger})\hat{x}^{\dagger}
- 2r^{-1}\sin^2f\partial_rf (\hat{x}_Ie_J^{\dagger}-\hat{x}_Je_I^{\dagger}) \notag \\
&= -D\Sigma_{IJ}^{(-)} + E\Theta_{IJ}\hat{x}^{\dagger} - F\Theta_{IJ},
\end{align}
where we have defined
$D=r^{-2}\sin^2f, E=2\left( r^{-2}\sin^2f - r^{-1}\sin f\cos
f\partial_rf \right), F=2r^{-1}\sin^2f\partial_rf$ and
$\Theta_{IJ}=\hat{x}_Ie_J^{\dagger}-\hat{x}_Je_I^{\dagger}$.
Here the matrices $\Sigma_{IJ}^{(-)}$ and $\Theta_{IJ}$ satisfy the following relations
\begin{align}
\Theta_{IJ}\hat{x}^{\dagger} &= -\hat{x}^{\dagger}\Theta_{IJ}, &
\Sigma_{IJ}^{(-)}\Theta_{IJ} &= 4\cdot6\hat{x}^{\dagger}, &
\Theta_{IJ}\Sigma_{IJ}^{(-)} &= -4\cdot6\hat{x}^{\dagger}, \notag \\
&& &\hspace{-100pt}\Theta_{IJ}^2 = -2\cdot6~\mathbf{1}_8, &
&\hspace{-100pt}\left( \Sigma_{IJ}^{(-)} \right)^2 = -4\cdot7\cdot6~\mathbf{1}_8.
\end{align}
The squares of the commutator $[R_I,R_J]$ is evaluated as 
\begin{align}
[R_I,R_J]^2 &= -24r^{-2}\sin^2f\left( 5r^{-2}\sin^2f+2(\partial_rf)^2 \right)\mathbf{1}_8.
\end{align}
Using this result, we can calculate the first term in
\eqref{eq:8dSkyrme} as 
\begin{align}
\left( [R_I,R_J]^2 \right)^2 = 16\cdot6^2r^{-4}\sin^4f\left( 25r^{-4}\sin^4f + 20r^{-2}\sin^2f(\partial_rf)^2 + 4(\partial_rf)^4 \right)\mathbf{1}_8.
\end{align}
Things get more involved when we calculate the second term.
We expand the second term in \eqref{eq:8dSkyrme} as
\begin{align}
&\hspace{-5pt}\left( [R_I,R_J][R_K,R_L] \right)^2
= D^4\Sigma^{(-)}_{IJ}\Sigma^{(-)}_{KL}\Sigma^{(-)}_{IJ}\Sigma^{(-)}_{KL} \notag \\
&\hspace{2pt}- D^3E\Bigl( \Sigma^{(-)}_{IJ}\Sigma^{(-)}_{KL}\Sigma^{(-)}_{IJ}\Theta_{KL}\hat{x}^{\dagger}+\Sigma^{(-)}_{IJ}\Sigma^{(-)}_{KL}\Theta_{IJ}\hat{x}^{\dagger}\Sigma^{(-)}_{KL}+\Sigma^{(-)}_{IJ}\Theta_{KL}\hat{x}^{\dagger}\Sigma^{(-)}_{IJ}\Sigma^{(-)}_{KL}+\Theta_{IJ}\hat{x}^{\dagger}\Sigma^{(-)}_{KL}\Sigma^{(-)}_{IJ}\Sigma^{(-)}_{KL} \Bigr) \notag \\
&\hspace{4pt}+ D^3F\Bigl( \Sigma^{(-)}_{IJ}\Sigma^{(-)}_{KL}\Sigma^{(-)}_{IJ}\Theta_{KL}+\Sigma^{(-)}_{IJ}\Sigma^{(-)}_{KL}\Theta_{IJ}\Sigma^{(-)}_{KL}+\Sigma^{(-)}_{IJ}\Theta_{KL}\Sigma^{(-)}_{IJ}\Sigma^{(-)}_{KL}+\Theta_{IJ}\Sigma^{(-)}_{KL}\Sigma^{(-)}_{IJ}\Sigma^{(-)}_{KL} \Bigr) \notag \\
&\hspace{6pt}+ D^2(E^2+F^2)\Bigl( \Sigma^{(-)}_{IJ}\Sigma^{(-)}_{KL}\Theta_{IJ}\Theta_{KL}+\Theta_{IJ}\Theta_{KL}\Sigma^{(-)}_{IJ}\Sigma^{(-)}_{KL} \Bigr) \notag \\
&\hspace{8pt}+ D^2E^2\Bigl( \Sigma^{(-)}_{IJ}\Theta_{KL}\hat{x}^{\dagger}\Sigma^{(-)}_{IJ}\Theta_{KL}\hat{x}^{\dagger}+\Sigma^{(-)}_{IJ}\Theta_{KL}\hat{x}^{\dagger}\Theta_{IJ}\hat{x}^{\dagger}\Sigma^{(-)}_{KL} \notag \\
&\hspace{100pt} +\Theta_{IJ}\hat{x}^{\dagger}\Sigma^{(-)}_{KL}\Sigma^{(-)}_{IJ}\Theta_{KL}\hat{x}^{\dagger}+\Theta_{IJ}\hat{x}^{\dagger}\Sigma^{(-)}_{KL}\Theta_{IJ}\hat{x}^{\dagger}\Sigma^{(-)}_{KL} \Bigr) \notag \\
&\hspace{10pt} - D^2EF\Bigl( \Sigma^{(-)}_{IJ}\Theta_{KL}\hat{x}^{\dagger}\Sigma^{(-)}_{IJ}\Theta_{KL}+\Sigma^{(-)}_{IJ}\Theta_{KL}\hat{x}^{\dagger}\Theta_{IJ}\Sigma^{(-)}_{KL}+\Theta_{IJ}\hat{x}^{\dagger}\Sigma^{(-)}_{KL}\Sigma^{(-)}_{IJ}\Theta_{KL}+\Theta_{IJ}\hat{x}^{\dagger}\Sigma^{(-)}_{KL}\Theta_{IJ}\Sigma^{(-)}_{KL}  \notag \\
&\hspace{22pt} + \Sigma^{(-)}_{IJ}\Theta_{KL}\Sigma^{(-)}_{IJ}\Theta_{KL}\hat{x}^{\dagger} + \Sigma^{(-)}_{IJ}\Theta_{KL}\Theta_{IJ}\hat{x}^{\dagger}\Sigma^{(-)}_{KL} + \Theta_{IJ}\Sigma^{(-)}_{KL}\Sigma^{(-)}_{IJ}\Theta_{KL}\hat{x}^{\dagger} + \Theta_{IJ}\Sigma^{(-)}_{KL}\Theta_{IJ}\hat{x}^{\dagger}\Sigma^{(-)}_{KL} \Bigr) \notag \\
&\hspace{12pt} -DE(E^2+F^2)\Bigl( \Sigma^{(-)}_{IJ}\Theta_{KL}\hat{x}^{\dagger}\Theta_{IJ}\Theta_{KL}+\Theta_{IJ}\hat{x}^{\dagger}\Sigma^{(-)}_{KL}\Theta_{IJ}\Theta_{KL} \notag \\
&\hspace{120pt} +\Theta_{IJ}\Theta_{KL}\Sigma^{(-)}_{IJ}\Theta_{KL}\hat{x}^{\dagger}+\Theta_{IJ}\Theta_{KL}\Theta_{IJ}\hat{x}^{\dagger}\Sigma^{(-)}_{KL} \Bigr) \notag \\
&\hspace{14pt} +D^2F^2\Bigl( \Sigma^{(-)}_{IJ}\Theta_{KL}\Sigma^{(-)}_{IJ}\Theta_{KL}+\Sigma^{(-)}_{IJ}\Theta_{KL}\Theta_{IJ}\Sigma^{(-)}_{KL}+\Theta_{IJ}\Sigma^{(-)}_{KL}\Sigma^{(-)}_{IJ}\Theta_{KL}+\Theta_{IJ}\Sigma^{(-)}_{KL}\Theta_{IJ}\Sigma^{(-)}_{KL} \Bigr) \notag \\
&\hspace{16pt} + DF(E^2+F^2)\Bigl( \Sigma^{(-)}_{IJ}\Theta_{KL}\Theta_{IJ}\Theta_{KL}+\Theta_{IJ}\Sigma^{(-)}_{KL}\Theta_{IJ}\Theta_{KL} \notag \\
&\hspace{120pt} +\Theta_{IJ}\Theta_{KL}\Sigma^{(-)}_{IJ}\Theta_{KL}+\Theta_{IJ}\Theta_{KL}\Theta_{IJ}\Sigma^{(-)}_{KL} \Bigr) \notag \\
&\hspace{18pt} + (E^2+F^2)^2\Theta_{IJ}\Theta_{KL}\Theta_{IJ}\Theta_{KL}.
\label{eq:2nd}
\end{align}
Here we have used the relation
$\Theta_{IJ}\hat{x}^{\dagger}\Theta_{KL}\hat{x}^{\dagger}=\Theta_{IJ}\Theta_{KL}$
and $\Theta_{IJ}\hat{x}^{\dagger}\Theta_{KL} +
\Theta_{IJ}\Theta_{KL}\hat{x}^{\dagger} = 0$.
We stress that terms that contain the odd number of $\hat{x}$ or
$\hat{x}^\dagger$ vanish under the trace of the matrices. 
Since we need the trace of \eqref{eq:2nd} in the energy functional, we
neglect these terms and never calculate them in the following.
Exploiting this fact, we are left with the terms that contain the even
number of $\hat{x}$:
\begin{align}
D^4~\text{term}&:&  \Sigma^{(-)}_{IJ}\Sigma^{(-)}_{KL}\Sigma^{(-)}_{IJ}\Sigma^{(-)}_{KL} &=& 1344&~\mathbf{1}_8, \notag \\
D^3E~\text{term}&:& \Sigma^{(-)}_{IJ}\Sigma^{(-)}_{KL}\Sigma^{(-)}_{IJ}\Theta_{KL}\hat{x}^{\dagger} +\dots+ \Theta_{IJ}\hat{x}^{\dagger}\Sigma^{(-)}_{KL}\Sigma^{(-)}_{IJ}\Sigma^{(-)}_{KL} &=& 4\cdot192&~\mathbf{1}_8, \notag \\
D^2(E^2+F^2)~\text{term}&:& \Sigma^{(-)}_{IJ}\Sigma^{(-)}_{KL}\Theta_{IJ}\Theta_{KL}+\Theta_{IJ}\Theta_{KL}\Sigma^{(-)}_{IJ}\Sigma^{(-)}_{KL} &=& -2\cdot384&~\mathbf{1}_8, \notag \\
D^2E^2~\text{term}&:& \Sigma^{(-)}_{IJ}\Theta_{KL}\hat{x}^{\dagger}\Sigma^{(-)}_{IJ}\Theta_{KL}\hat{x}^{\dagger} + \Theta_{IJ}\hat{x}^{\dagger}\Sigma^{(-)}_{KL}\Theta_{IJ}\hat{x}^{\dagger}\Sigma^{(-)}_{KL} &=& 2\cdot96&~\mathbf{1}_8, \notag \\
	&& \Sigma^{(-)}_{IJ}\Theta_{KL}\hat{x}^{\dagger}\Theta_{IJ}\hat{x}^{\dagger}\Sigma^{(-)}_{KL} + \Theta_{IJ}\hat{x}^{\dagger}\Sigma^{(-)}_{KL}\Sigma^{(-)}_{IJ}\Theta_{KL}\hat{x}^{\dagger} &=& -2\cdot384&~\mathbf{1}_8, \notag \\
DE(E^2+F^2)~\text{term}&:& \Sigma^{(-)}_{IJ}\Theta_{KL}\hat{x}^{\dagger}\Theta_{IJ}\Theta_{KL} +\dots+ \Theta_{IJ}\Theta_{KL}\Theta_{IJ}\hat{x}^{\dagger}\Sigma^{(-)}_{KL} &=& -4\cdot192&~\mathbf{1}_8, \notag \\
D^2F^2~\text{term}&:& \Sigma^{(-)}_{IJ}\Theta_{KL}\Sigma^{(-)}_{IJ}\Theta_{KL} + \Theta_{IJ}\Sigma^{(-)}_{KL}\Theta_{IJ}\Sigma^{(-)}_{KL} &=& 2\cdot864&~\mathbf{1}_8, \notag \\
	&& \Sigma^{(-)}_{IJ}\Theta_{KL}\Theta_{IJ}\Sigma^{(-)}_{KL} + \Theta_{IJ}\Sigma^{(-)}_{KL}\Sigma^{(-)}_{IJ}\Theta_{KL} &=& -2\cdot384&~\mathbf{1}_8, \notag \\
(E^2+F^2)^2~\text{term}&:& \Theta_{IJ}\Theta_{KL}\Theta_{IJ}\Theta_{KL} &=& -96&~\mathbf{1}_8.
\end{align}
With this result at hand, we find that the second term in
\eqref{eq:8dSkyrme} becomes
\begin{equation}
\text{Tr}\left( [R_I,R_J][R_K,R_l] \right)^2 =  1536r^{-4}\sin^4f\Bigl( -5r^{-4}\sin^4f + 20r^{-2}\sin^2f(\partial_rf)^2 - 8(\partial_rf)^4 \Bigr).
\end{equation}
We can calculate the other terms by same method.
After the calculations, the results are 
\begin{align}
\text{Tr}\left( [R_I,R_J]^2 \right)^2 &= 4608r^{-4}\sin^4f\left( 25r^{-4}\sin^4f + 20r^{-2}\sin^2f(\partial_rf)^2 + 4(\partial_rf)^4 \right), \notag \\
\text{Tr}\left( [R_I,R_J][R_K,R_l] \right)^2 &=  1536r^{-4}\sin^4f\Bigl( -5r^{-4}\sin^4f + 20r^{-2}\sin^2f(\partial_rf)^2 - 8(\partial_rf)^4 \Bigr), \notag \\
\text{Tr}[R_I,R_J][R_K,R_l][R_I,R_K][R_J,R_l] &=  768r^{-4}\sin^4f\left( -55r^{-4}\sin^4f - 80r^{-2}\sin^2f(\partial_rf)^2 + 2(\partial_rf)^4 \right), \notag \\
\text{Tr}\left( [R_I,R_J] \right)^2R_K^2 &= 192r^{-2}\sin^2f\left( 30r^{-4}\sin^4f + 17r^{-2}\sin^2f(\partial_rf)^2 + 2(\partial_rf)^4 \right), \notag \\
\text{Tr}\left( [R_I,R_J]R_K \right)^2 &=  192r^{-2}\sin^2f\left( 10r^{-4}\sin^4f + 13r^{-2}\sin^2f(\partial_rf)^2 - 2(\partial_rf)^4 \right), \notag \\
\text{Tr}[R_I,R_J]R_K[R_I,R_K]R_J &= 192r^{-2}\sin^2f\left( -25r^{-4}\sin^4f - 16r^{-2}\sin^2(\partial_rf)^2 - (\partial_rf)^4 \right), \notag \\
\text{Tr}[R_I,R_J][R_K,R_I]R_JR_K &= 192r^{-2}\sin^2f\left( 15r^{-4}\sin^4f + 14r^{-2}\sin^2f(\partial_rf)^2 - (\partial_rf)^4 \right), \notag \\
\text{Tr}[R_I,R_J]R_I[R_K,R_J]R_K &= 192r^{-2}\sin^2f\left( -25r^{-4}\sin^4f - 16r^{-2}\sin^2(\partial_rf)^2 - (\partial_rf)^4 \right).
\end{align}

Collecting everything altogether, we finally obtain
\begin{align}
&\text{Tr}\biggl[ c_2\left( [R_I,R_J][R_I,R_J] \right)^2 + c_2\left( [R_I,R_J][R_K,R_L] \right)^2 - 4c_2[R_I,R_J][R_K,R_L][R_I,R_K][R_J,R_L] \notag \\
&\hspace{30pt} + 4c_1\left( [R_I,R_J] \right)^2R_K^2 + 4c_1\left( [R_I,R_J]R_K \right)^2 - 4c_1[R_I,R_J]R_K[R_I,R_K]R_J \notag \\
&\hspace{120pt}+ 8c_1[R_I,R_J][R_K,R_I]R_JR_K - 4c_1[R_I,R_J]R_I[R_K,R_J]R_K \biggr] \notag \\
&= 23040\left( 3c_1r^{-4}\sin^4f(\partial_rf)^2 + 4r^{-6}\sin^6f\left( 4c_2(\partial_rf)^2 + c_1 \right) +  12c_2r^{-8}\sin^8f \right).
\end{align}
Taking $c_1 = c_2 = 1$ and introducing the overall factor
$\frac{16}{15}\pi^3r^6$, we obtain the energy functional
\eqref{eq:8dSkymre_action_hedgehog}.
Here we have taken into account the factor that comes from the six-dimensional spherical integration:
\begin{equation}
\int_{S^6}d\Omega_6 = \frac{16}{15}\pi^3r^6,
\end{equation}
where $S^6$ is the six-dimensional spherical surface and $d\Omega_6$ is
the integral element of the six-dimensional sphere.

\end{appendix}

\end{document}